\documentclass[twocolumn,preprintnumbers,amsmath,amssymb]{revtex4}
\usepackage{amsfonts}
\usepackage{graphicx}

\begin{document}

\title{Magnetic Metamaterials at Telecommunication and Visible Frequencies}
\author{C. Enkrich$^1$, M. Wegener$^{1,2}$, S. Linden$^2$, S. Burger$^3$, L. Zschiedrich$^3$, F. Schmidt$^3$, J.\,F. Zhou$^4$, Th. Koschny$^4$, and C.\,M. Soukoulis$^{4,5}$}
\affiliation{$^1$ Institut f\"ur Angewandte Physik, Universit\"at
Karlsruhe (TH), D-76131 Karlsruhe, Germany} \affiliation{$^2$ Institut
f\"ur Nanotechnologie, Forschungszentrum Karlsruhe in der
Helmholtz-Gemeinschaft, D-76021 Karlsruhe, Germany} \affiliation{$^3$ Zuse
Institute Berlin , Takustra{\ss}e 7, D-14195 Berlin, Germany and \\ DFG
Forschungszentrum {\sc Matheon}, Stra{\ss}e des 17. Juni 136, 10623
Berlin, Germany } \affiliation{$^4$ Ames Laboratory and Department of
Physics and Astronomy, Iowa State University, Ames, Iowa 50011, U.S.A.}
\affiliation{$^5$ Institute of Electronic Structure and Laser, FORTH and
Department of Materials Science and Technology, 71110 Heraklion, Crete,
Greece}
\date{\today}

\begin{abstract}
Arrays of gold split rings with a 50-nm minimum feature size and with an
$LC$ resonance at 200 THz frequency (1.5 $\rm \mu m$ wavelength) are
fabricated. For normal-incidence conditions, they exhibit a pronounced
fundamental magnetic mode, arising from a coupling via the electric
component of the incident light. For oblique incidence, a coupling via the
magnetic component is demonstrated as well. Moreover, we identify a novel
higher-order magnetic resonance at around 370\,THz (800 nm wavelength)
that evolves out of the Mie resonance for oblique incidence. Comparison
with theory delivers good agreement and also shows that the structures
allow for a negative magnetic permeability.

\copyright{2005 The American Physical Society}
\end{abstract}


\maketitle

In usual crystals, the atoms are arranged in a periodic fashion with
lattice constants less than 1 nm. This is orders of magnitude smaller than
the wavelength of visible light. Thus, the light experiences an effective
homogeneous material; it does not ``see'' the underlying periodicity
(apart from crystal symmetries). Microscopically, the light excites
electric dipoles that reradiate with a certain retardation, slowing down
the phase velocity of light in the material by a factor called the index
of refraction. Metamaterials are artificial periodic structures with
``lattice constants'' that are still smaller than the wavelength of light.
Again, the light field ``sees'' an effective homogeneous material. The
``atoms,'' however, are not real atoms but are rather artificial
nanostructures composed of many atoms. This allows for tailoring their
properties in a way not possible with normal atoms. Indeed, Pendry
\cite{Pendry1999} showed that a combination of ``magnetic atoms'' and
``electric atoms'' (i.e., split-rings and metallic wires) with negative
permeability $\mu$ and permittivity $\epsilon$, respectively, can lead to
materials with a negative index of refraction $n$ \cite{Veselago1968}.
These materials open a whole new chapter of photonics connected with novel
concepts and potential applications \cite{Pendry2000,Smith2004}.

The main technological challenge is to obtain a negative permeability $\mu
< 0$ at telecom or visible frequencies, which does not occur in natural
materials. Starting with first demonstrations in the microwave regime
\cite{Shelby2001}, the achieved magnetic resonance frequencies have
increased by more than 4 orders of magnitude over the last four years
\cite{Shelby2001,Yen2004,Linden2004,Zhang2005,Moser2005,Katsarakis2005},
reaching a record of 100\,THz (3 $\rm \mu m$ wavelength) in November 2004
\cite{Linden2004}.

So far, the ``magnetic atom'' of choice has been the split ring resonator
(SRR), in essence just a small {\it LC} circuit consisting of an
inductance {\it L} and a capacitance {\it C}. The resonance frequency of
this {\it LC} circuit scales inversely with its size, provided the
frequencies are significantly below the metal plasma frequency. Near the
resonance, the current in the inductance can lead to a magnetic field
opposing the external magnetic field of the light, hence enabling $\mu <
0$.

The design used here closely follows our recent approach based on arrays
of single SRR \cite{Linden2004,Katsarakis2004}. The structures are
fabricated using standard electron-beam lithography on a 1 mm thick glass
substrate coated with a 5 nm thin film of indium-tin-oxide (ITO), in order
to avoid charging effects of the poly(methyl methacrylate) resist layer
(PMMA 950k) during the exposure. The gold film thickness is 30\,nm. To
increase the resonance frequency at a given minimum feature size and to
simplify the nanofabrication, we almost eliminate the tiny upper arms of
the SRR, leading to more ``{\it U}''-shaped structures (see Fig.\,1 (a)).
Intuitively, these {\it U}s correspond to $\frac{3}{4}$ of one winding of
a magnetic coil with inductance {\it L}. The ends of the {\it U}-shaped
wire form the capacitance {\it C}. We employ $(100\,\rm \mu m)^2$ periodic
quadratic arrays of such SRR with the dimensions apparent from the
electron micrographs in Fig.\,1.

\begin{figure}
\centerline{\includegraphics[width=6cm,keepaspectratio]{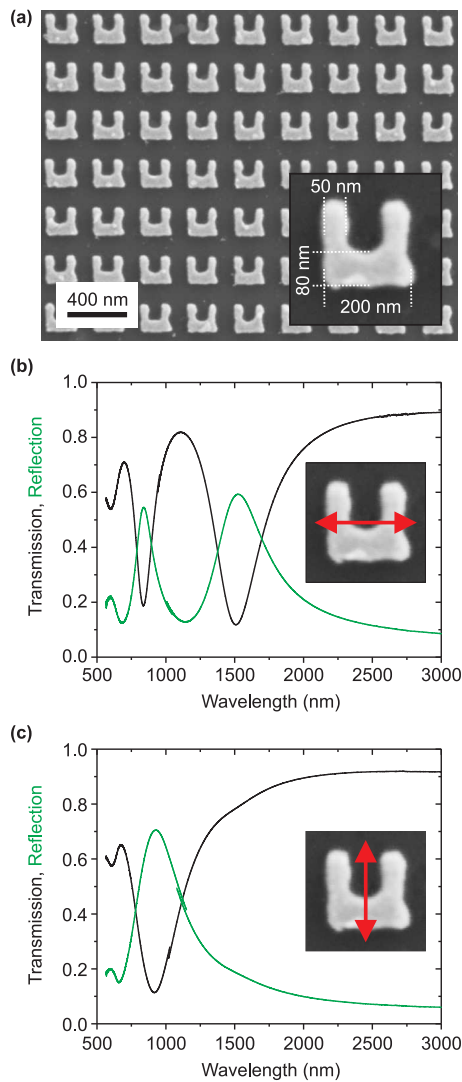}}
\caption{\small Electron micrograph of a split-ring array with a total
size of $(100\,\rm \mu m)^2$. The lower right-hand side inset shows  the
dimensions of an individual split ring. The corresponding measured
normal-incidence transmission and  reflection spectra for horizontal and
vertical polarization are shown in (b) and (c), respectively.  For (b),
one can couple to the fundamental magnetic mode at 1.5 $\rm \mu m$
wavelength via the electric-field component of  the incident light; for
(c), one cannot.} \label{Fig1}
\end{figure}

Figure 1 connects to our previous work \cite{Linden2004} and shows spectra
taken under normal incidence with a commercial Fourier-transform
microscope spectrometer \cite{online_material}. If the incident light is
polarized horizontally, the electric field can couple to the capacitance
of the SRR and induce a circulating current in the coil leading to a
magnetic-dipole moment normal to the SRR plane. Note that this resonance
at 1.5 $\rm \mu m$ wavelength is yet more pronounced than in our previous
work \cite{Linden2004} at 3 $\rm \mu m$, due mainly to the increased ratio
between thickness and lateral size of the SRR. The magnetic resonance
disappears for vertical incident polarization (Fig.\,1(c)), leaving behind
only the Mie resonance of the SRR around 950 nm wavelength. As this
resonance will become important below as well, we briefly recapitulate its
physics. Here the electric field of the light leads to a charge
accumulation at the surfaces of the vertical SRR arms, resulting in a
depolarization field. Depending on the permittivity of the metal, hence
depending on the frequency of light, this depolarization field can enhance
or suppress the external electric field. (We also observe a weaker
short-wavelength Mie resonance around 600-nm in Fig.\,1, which is due to
the depolarization field of the short axis, i.e., the width of the SRR
arms.) Notably, the fundamental Mie resonance of our SRR changes in
spectral position and width between the two different polarization
configurations. This can be understood as follows: For horizontal incident
polarization and for the frequencies of interest here, only the
fundamental Mie resonance of the SRR bottom arm is excited. For vertical
polarization, the two similarly shaped vertical SRR arms contribute. The
latter are coupled via the SRR's bottom arm (and via the radiation field).
As usual, the coupling of two degenerate modes leads to an avoided
crossing with two new effective oscillation modes, a symmetric and an
antisymmetric one, which are frequency down-shifted and up-shifted as
compared to the uncoupled resonances, respectively. The antisymmetric mode
cannot be excited at all for normal incidence as it has zero effective
electric-dipole moment. The redshifted symmetric mode can be excited. It
even has a larger effective electric-dipole moment than a single arm.

The optical response of SRR is not only polarization-dependent but also
highly anisotropic. Thus, we have performed transmission experiments under
oblique incidence (Fig.\,2) using a dedicated home-built setup
\cite{online_material}. Compared with the Fourier-transform microscope
spectrometer used above, this setup has improved polarization optics by
using Glan-Thomson polarizers and a smaller effective opening angle ($\pm
7$ degrees). It still allows one to investigate a sample of size
(100\,$\mu$m)$^2$ with a large angle of incidence with respect to the
surface normal over a broad spectral range. In Fig.\,2(a), the electric
component of the incident light can not couple to the {\it LC} circuit
resonance for any angle. With increasing angle, however, the magnetic
field acquires a component normal to the SRR plane. This component can
induce a circulating electric current in the SRR coil via the induction
law (see right-hand side inset in Fig.\,3(a)). This current again leads to
a magnetic-dipole moment normal to the SRR plane, which can counteract the
external magnetic field. The magnitude of this resonance (highlighted by
the blue area around 1.5 $\rm \mu m$ wavelength) is indeed consistent with
theory (see below) and leads to an effective negative magnetic
permeability for propagation in the SRR plane and for a stack of SRR
layers rather than just one layer considered here. This aspect has been
verified explicitly by retrieving the effective permittivity and
permeability from the calculated transmission and reflection spectra
\cite{retrieval_Smith,retrieval_Koschny}. The shape of the retrieved
magnetic permeability closely resembles that published in our previous
work \cite{Linden2004} at 3 $\rm \mu m$ wavelength. It exhibits a negative
permeability with a minimum value of $\mu=-0.25$ at 1.67 $\rm \mu m$
wavelength \cite{online_material}. This value could be further improved by
increasing the number of SRR per area (compare Fig.\,1(a)), hence
increasing the ``oscillator strength'' of the magnetic resonance.

\begin{figure}
\centerline{\includegraphics[width=6.2cm,keepaspectratio]{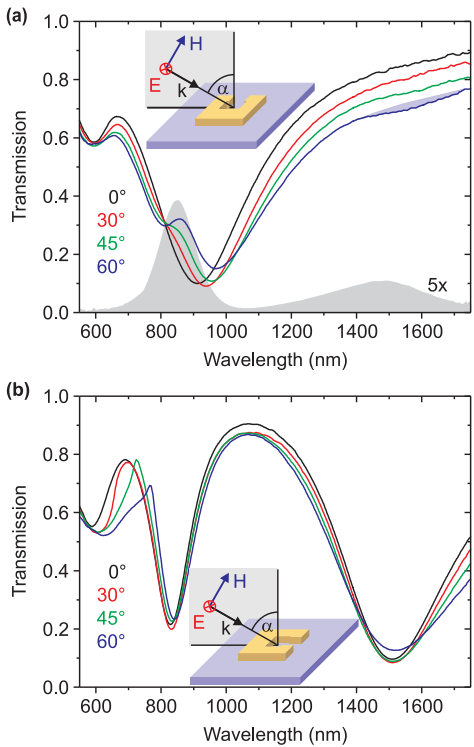}}
\caption{\small Measured transmission spectra taken for oblique incidence
for the configurations shown as insets (where $\alpha  = 60^\circ$). In
(a), coupling to the fundamental magnetic mode at 1.5 $\rm \mu m$
wavelength is only possible via the magnetic-field component of the
incident light; for (b), both electric and magnetic field can couple. Note
the small but significant feature in (a) for 60$^\circ$ around 1.5-$\rm
\mu m$ wavelength. The lower gray area in (a) is the transmission into the
linear polarization orthogonal to the incident one for $\alpha  =
60^\circ$. We argue that this observable can be viewed as a fingerprint of
magnetic resonances under these conditions.} \label{Fig2}
\end{figure}

Another striking feature of the spectra in Fig.\,2(a) is that the 950 nm
wavelength Mie resonance at normal incidence splits into two resonances
for oblique incidence. This aspect can be understood continuing along the
lines of our above intuitive discussion: For finite angles of incidence,
the phase fronts of the electric field are tilted with respect to the SRR
plane. Thus, the vertical SRR arms are excited with a small but finite
time delay, equivalent to a finite phase shift. This shift allows coupling
to the antisymmetric mode of the coupled system of the two vertical arms
as well. In one half cycle of light, one gets a positive charge at the
lower left-hand side corner of the SRR and a negative charge at the lower
right-hand side corner, resulting in a compensating current in the
horizontal bottom arm. Characteristic snapshots of the current
distributions in the SRR are schematically shown as insets in Fig.\,3(a).
Altogether, we get a part of a circulating current, leading to a
magnetic-dipole moment. This type of magnetic resonance has not been
observed before. Its spectral position of 800\,nm at 60$^\circ$ angle in
Fig.\,2(a) is just within the visible regime -- for the first time, one
can literally see a magnetic resonance.

According to our reasoning for oblique incidence (e.g., 60$^\circ$), we
expect a circulating current component for wavelengths near the two
magnetic resonances at $1.5\,\rm \mu m$ and 800\,nm, respectively. Any
circulating current is evidently connected with a current in the
horizontal bottom arm of the SRR. According to the usual laws of a Hertz
dipole, the corresponding charge oscillation in the bottom arm can radiate
into the forward direction with an electric field component orthogonal to
the incident polarization. In other words, for oblique incidence, the
fingerprint of the magnetic resonances is a rotation of polarization. Such
rotation is indeed unambiguously observed in our experiments (see gray
area in Fig.\,2(a)), strongly supporting our above interpretation.

\begin{figure}
\centerline{\includegraphics[width=6.2cm,keepaspectratio]{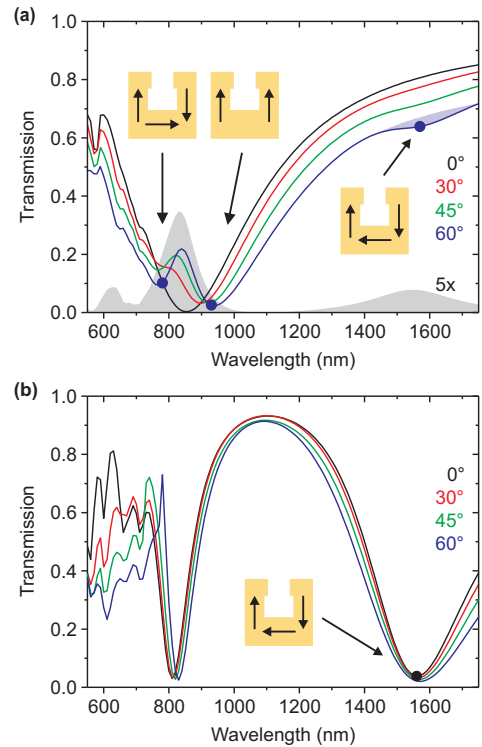}}
\caption{\small Calculated transmission spectra for different angles on
the same scale as the experiment (Fig.\,2). The polarization geometries in
(a) and (b) are identical to those in Figs. 2(a) and 2(b), respectively.
The insets schematically show the current distributions at the frequencies
and angles marked by the dots.} \label{Fig3}
\end{figure}

Figure 2(b) shows spectra for a second polarization configuration in which
both the electric and the magnetic component of the light field can couple
to the SRR for oblique incidence. The same configuration was used in
Ref.\,\cite{Yen2004}. Our results immediately show that coupling to the
fundamental magnetic resonance at 1.5 $\rm \mu m$ wavelength is not
mediated by the magnetic field alone here. The two further independent
polarization and angle configurations are consistent with our above
interpretation but do not reveal any additional resonances
\cite{online_material}.

In order to further strengthen our above interpretation of the different
resonances, we compare the measured spectra with theory. It turns out that
numerical calculation of spectra for oblique incidence is much more of a
challenge than for normal incidence or for propagation in the plane of the
SRR array. Thus, we have not only followed our previous finite-difference
time-domain approach \cite{Linden2004} but also an advanced
frequency-domain based finite-element approach. We discretize the geometry
of a unit cell ($315\,{\rm nm} \times 330\,{\rm nm} \times 60\,{\rm nm}$)
including the SRR and the surrounding layered media with an unstructured
mesh of tetrahedra. The coarse mesh is automatically refined to a fine
mesh consisting about 6000 tetrahedra. Bloch-periodic boundary conditions
are applied in the $x$ and $y$ directions \cite{Burger2004} with lattice
constants of $a_x=315\,\rm nm$ and $a_y=330\,\rm nm$, respectively. In the
$\pm z$ direction we apply transparent boundary conditions
\cite{Zschiedrich2005}. To avoid dealing with the complex SRR shape
apparent from Fig.\,1, the three SRR arms are approximated as rectangles
with a width of 65\,nm for the two vertical arms and 90\,nm for the bottom
arm. These values have been fine-tuned to fit the calculated to the
measured resonance positions. The SRR side length of 200\,nm and the gold
thickness of 30\,nm are identical to the values already quoted above. For
the permittivity of gold we assume a wavelength dependence according to
the Drude model, with a plasma frequency of $\omega_{\rm pl} = 1.367
\times 10^{16}\,\rm s^{-1}$ and a collision frequency of $\omega_{\rm c} =
6.478 \times 10^{13}\,\rm s^{-1}$. The permittivity of the ITO layer
(glass substrate) is taken as 3.8 (2.25). We discretize Maxwell's
equations using vectorial finite elements (Whitney elements) of second
polynomial order. The resulting sparse matrix equation (with about
130\,000 unknowns corresponding to the solution on the fine mesh) is
solved on a standard personal computer by either standard linear algebra
decomposition techniques or multigrid algorithms, depending on the problem
size.

Figure 3 exhibits calculated spectra for different angles of incidence.
The graphical representation is identical to that of the experiment (see
Fig.\,2). Obviously, the qualitative agreement between experiment and
theory is excellent. Especially the spectral positions as well as the
magnitudes of the different resonances and the rotation of the
polarization (gray area at the bottom of Fig.\,3(a)) are well reproduced.
For normal incidence, the spectra calculated with the finite-element
frequency-domain approach agree well with those obtained from the
finite-difference time-domain simulations, which are used for the
retrieval procedure (see above). The schematic insets shown in Fig.\,3
have been derived from the movies \cite{online_material} of the calculated
field distributions and correspond to the qualitative discussion given
above.

In conclusion, normal optical materials exhibit only an {\it
electric-dipole} response. Here we have demonstrated pronounced {\it
magnetic-dipole} modes at 1.5 $\rm \mu m$ and at 800 nm wavelength,
respectively. The one at 1.5 $\rm \mu m$ is the ``usual'' $LC$ resonance
of the split ring resonators and would lead to a negative magnetic
permeability indeed. The one at 800 nm wavelength is a higher-order
magnetic resonance, which is identified here for the first time. On the
one hand, the nanofabrication of magnetic metamaterials at these
frequencies requires much more of a technological effort than at microwave
frequencies.  On the other hand, linear optical spectroscopy can be
performed much more conveniently and in a more controlled fashion.
Moreover, the availability of lasers in this spectral regime enables
future nonlinear optical experiments.

We acknowledge the support by the DFG-Center for Functional Nanostructures
within subproject A\,1.5, by DFG project We-1497/9-1, by DFG priority
programme SPP\,1113,  and by the BMBF within project 13N8252. The research
of C.\,M.\,S. is further supported by the Alexander von Humboldt
senior-scientist award 2002, by Ames Laboratory (Contract No.
W-7405-Eng-82), EU FET project DALHM and DARPA (Contract No. MDA
972-01-2-0016).

{\it Note added in proof.} -- Similar U-shaped SRR have recently also been
discussed theoretically in \cite{wirepair3}.

\end{document}